\documentclass[aps,prd,showpacs,preprintnumbers,nofootinbib]{revtex4}
\usepackage[dvips]{graphicx}
\usepackage{bm,latexsym,amsmath,amssymb,amsfonts}
\newcommand*{\car}{{\cal R}}
\begin{document}

\title{
A Vaidya-type radiating solution in Einstein-Gauss-Bonnet gravity
and its application to braneworld
}

\author{Tsutomu Kobayashi}
\email{tsutomu@tap.scphys.kyoto-u.ac.jp}

\affiliation{
Department of Physics, Kyoto University, Kyoto 606-8502, Japan 
}


\begin{abstract}
We consider a Vaidya-type radiating spacetime
in Einstein gravity with the Gauss-Bonnet combination of quadratic curvature terms.
Simply generalizing the known static black hole solutions
in Einstein-Gauss-Bonnet gravity,
we present an exact solution in arbitrary dimensions
with the energy-momentum tensor given by a null fluid form.
As an application, we derive an evolution equation for
the ``dark radiation'' in the Gauss-Bonnet braneworld.
\end{abstract}


\preprint{KUNS-1962}

\maketitle

\section{Introduction}

Higher order curvature terms naturally arise
in various contexts.
Among theories of gravity that involve higher derivative curvature terms,
so-called Einstein-Gauss-Bonnet gravity and
its Lovelock generalization~\cite{Deruelle:2003ck}
are of particular interest because of their special features.
The field equations of the Einstein-Gauss-Bonnet (and the Lovelock) theory
contain derivatives of the metric of order no higher than the second,
and the linearized theory about flat spacetime is free of ghosts.
The Gauss-Bonnet term is given by
a combination of quadratic curvature terms as
\begin{eqnarray}
{\cal L}_{GB}=\car^2-4\car_{cd}\car^{cd}+\car_{cdef}\car^{cdef},
\label{GB-Lagrangian}
\end{eqnarray}
which is relevant in five dimensions or higher,
and reduces to a total derivative in four dimensions.
This Lagrangian appears in the low energy effective action
of the heterotic string theory
and in the loop correction due to quantum fields in curved backgrounds.
Very recently the Gauss-Bonnet correction is often discussed
in the braneworld
context~\cite{Nojiri:2000gv, gaussbonnet, Neupane:2001st, NandS, Charmousis:2002rc}
as an attempt to extend
the well-known Randall-Sundrum model~\cite{RS} in a natural way.

In Einstein-Gauss-Bonnet gravity
less number of exact solutions has been known so far.
Static and spherically symmetric black hole solutions with or without
a cosmological constant,
as well as a topological black hole
in an anti-de Sitter (AdS) spacetime
were obtained~\cite{Boulware:1985wk, Crisostomo:2000bb, Aros:2000ij, Cai:2003gr, Charmousis:2002rc, Cai:2001dz}.
They are the Gauss-Bonnet generalization
of the Schwarzschild and (A)dS-Schwarzschild solutions.
In this paper we present another new example of exact solutions:
a Gauss-Bonnet generalization of the Vaidya solution,
which in Einstein gravity originally describes
a spacetime outside a radiating ``star''
with the energy-momentum tensor of a null fluid~\cite{Vaidya}.
In the recent literature the Vaidya-type solution
in an AdS spacetime is frequently used to investigate
consequences of the braneworld emitting
gravitons into the extra
dimension~\cite{Langlois:2002ke, Leeper:2003dd, Gergely:2004zq, Vernon:2004ae, Jennings:2004wz}.
Using our new Vaidya-type solution,
we construct a toy model for a radiating Gauss-Bonnet braneworld,
giving an evolution equation for the ``dark radiation'' parameter.

\section{Einstein-Gauss-Bonnet gravity}

We begin with a brief review of Einstein-Gauss-Bonnet gravity
and static black hole solutions in it.
The gravitational part of the action that we consider is given by,
\begin{eqnarray}
{\cal S}=\frac{1}{2\kappa^2}\int d^Dx \sqrt{-g}
\left[\car - 2\Lambda + \alpha {\cal L}_{GB}
\right],
\end{eqnarray}
where $\kappa$ is the $D$-dimensional gravitational constant and
$\Lambda$ is a cosmological constant.
The Gauss-Bonnet Lagrangian ${\cal L}_{GB}$ is defined by Eq.~(\ref{GB-Lagrangian}),
and thus the parameter $\alpha$ has dimension of (length)$^2$.
We assume that $\alpha$ is positive.
We also assume that $D\geq 5$
because ${\cal L}_{GB}$ is just a total derivative in four dimensions.
From the action we obtain the following field equations:
\begin{eqnarray}
{\cal G}_{ab}=\kappa^2{\cal T}_{ab}
-\Lambda g_{ab}+\frac{\alpha}{2}{\cal H}_{ab},
\label{field_eq}
\end{eqnarray}
where ${\cal G}_{ab}$ is the Einstein tensor,
${\cal T}_{ab}$ is the energy-momentum tensor
of matter fields, and
\begin{eqnarray}
{\cal H}_{ab}:=
{\cal L}_{GB} g_{ab}
-4(\car\car_{ab}-2\car_{ac}\car^c_{~b}
-2\car_{acbd}\car^{cd}+\car_{acde}\car_b^{~cde}),
\end{eqnarray}
is the Lanczos tensor.

The maximally symmetric spacetimes,
\begin{eqnarray}
\car_{abcd}={\cal K}\left(g_{ac}g_{bd}-g_{ad}g_{bc}\right),
\end{eqnarray}
with the curvature scale
\begin{eqnarray}
&&{\cal K}:=-\frac{1}{2\tilde\alpha}\left[1\pm
\sqrt{1+\frac{8\tilde\alpha\Lambda}{(D-1)(D-2)}} \right],\\
&&\quad\tilde \alpha:=(D-3)(D-4)\alpha,
\end{eqnarray}
solve the field equations~(\ref{field_eq}).
Only in the ``$-$'' branch
we have ${\cal K}=2\Lambda/[(D-1)(D-2)]$ (the Einstein value)
in the limit of $\alpha\to 0$.
Although the ``$+$'' branch seems physically less important,
it is interesting to note that
even when $\Lambda =0$
we have ${\cal K}<0$, namely, AdS spacetime.

Einstein-Gauss-Bonnet gravity admits
static black hole solutions as a generalization of Schwarzschild
and (A)dS-Schwarzschild spacetime.
They are given by the metric of
the following form~\cite{Boulware:1985wk, Cai:2003gr, Charmousis:2002rc, Cai:2001dz}:
\begin{eqnarray}
ds^2=-f(r)dt^2+\frac{1}{f(r)}dr^2+r^2\gamma_{ij}dx^idx^j,
\label{BH_metric}
\end{eqnarray}
where $\gamma_{ij}$ is a metric of the constant curvature $(D-2)$-space
parameterized by $k=0$ (flat space), $+1$ (sphere), $-1$ (hyperboloid),
and
\begin{eqnarray}
f(r)=k+
\frac{r^2}{2\tilde \alpha}
\left[1 \pm \sqrt{1+\frac{8\tilde \alpha\Lambda}{(D-1)(D-2)}
+\frac{4\tilde \alpha M}{r^{D-1}}}
\right],
\end{eqnarray}
with $M$ being gravitational mass
(up to a constant factor which is not important here).
Here we assume that the expression in the square root
is positive.
In addition to the usual central singularity at $r=0$,
it can be shown that the Kretschmann scalar $\car_{abcd}\car^{abcd}$
diverges at the place where
the square root vanishes,
and thus there is potentially another singularity.
Again, in the $\alpha\to 0$ limit the ``$+$'' branch is ill-behaved,
while in the ``$-$'' branch we recover the Einstein result,
\begin{eqnarray}
f(r)\to k-\frac{2\Lambda}{(D-1)(D-2)}r^2-\frac{M}{r^{D-3}}.
\end{eqnarray}
Unless the cosmological constant is negative,
the constant curvature $(D-2)$-space must respect
a spherical symmetry $(k=1)$ because otherwise
one has $f(r)$ with the wrong signature for the ``$-$'' branch.

For the ``$+$'' branch the asymptotic behavior of the metric is
\begin{eqnarray*}
f(r) \approx k-{\cal K}r^2 +\frac{M}{r^{D-3}}
\left[\sqrt{1+\frac{8\tilde \alpha\Lambda}{(D-1)(D-2)}}\right]^{-1},
\end{eqnarray*}
from which one may conclude that this solution
has \textit{negative} mass for the ``$+$'' branch.
However, the properly defined mass with the Gauss-Bonnet
effects taken into account~\cite{Deser:2002jk, Deruelle:2003ps, Petrov:2005qt}
is in fact \textit{positive} for both branches.



For later convenience we rewrite the metric~(\ref{BH_metric})
by using a null coordinate defined by
\begin{eqnarray}
v:=t+\epsilon\int\frac{dr}{f(r)},
\end{eqnarray}
as
\begin{eqnarray}
ds^2=-f(r)dv^2+2\epsilon~dvdr+r^2\gamma_{ij}dx^idx^j,
\label{static_null}
\end{eqnarray}
where the trajectory of $v=$ constant corresponds to
the ingoing radial light rays for $\epsilon=+1$
and the outgoing ones for $\epsilon=-1$.

\section{A radiating solution in Einstein-Gauss-Bonnet gravity}

The original Vaidya solution in Einstein gravity describes
the spacetime surrounding a star
idealized as a radiating sphere~\cite{Vaidya}.
It is easy to generalize the solution in arbitrary spacetime dimensions 
with or without a cosmological constant, and
the metric is given by\footnote{
This expression includes the $D=4$ case
because it is a solution in Einstein gravity.}
\begin{eqnarray}
ds^2=-\left[k-\frac{2\Lambda}{(D-1)(D-2)}r^2-\frac{M(v)}{r^{D-3}}
\right]dv^2
+2\epsilon ~dvdr+r^2\gamma_{ij}dx^idx^j,
\label{AdS-Vaidya}
\end{eqnarray}
where $v$ is the ingoing (outgoing) null coordinate for $\epsilon=+1~(-1)$,
and the mass parameter $M$ is now a function of $v$.
From the Einstein equations
${\cal G}_a^{~b}+\Lambda\delta_a^{~b}=\kappa^2{\cal T}_a^{~b}$,
we see that the only nonvanishing component of
the energy-momentum tensor is
\begin{eqnarray}
\kappa^2{\cal T}_v^{~r}=\frac{D-2}{2r^{D-2}}\frac{dM}{dv},
\label{T_ab}
\end{eqnarray}
and thus it must be of the form
\begin{eqnarray}
{\cal T}_{ab}=\sigma(r,v)k_ak_b,
\label{null_fluid}
\end{eqnarray}
where $k^a$ is an ingoing (outgoing) null vector for $\epsilon=+1~(-1)$.
A fluid whose energy-momentum tensor is given by~(\ref{null_fluid})
is referred to as a \textit{null fluid}.

Let us consider such a radiating spacetime filled with a null fluid
in the framework of Einstein-Gauss-Bonnet gravity.
Starting from the static black hole solution
in the Einstein-Gauss-Bonnet gravity~(\ref{static_null}),
we can obtain a generalized Vaidya metric
simply by allowing the mass parameter to depend on
the null coordinate:
\begin{eqnarray}
M\to M(v).
\end{eqnarray}
Namely, the Vaidya-type radiating solution in Einstein-Gauss-Bonnet gravity
is given by
\begin{eqnarray}
ds^2=-f(r,v)dv^2+2\epsilon~drdv+r^2\gamma_{ij} dx^idx^j,
\label{Vaidya-GB}
\end{eqnarray}
where
\begin{eqnarray}
f(r, v)=k+
\frac{r^2}{2\tilde \alpha}
\left[1 \pm \sqrt{1+\frac{8\tilde \alpha\Lambda}{(D-1)(D-2)}
+\frac{4\tilde \alpha M(v)}{r^{D-1}}}
\right].
\end{eqnarray}
This metric is indeed a solution of the field equations~(\ref{field_eq})
with the only nonvanishing component
of the energy-momentum tensor
$\kappa^2{\cal T}_v^{~r}=[(D-2)/2r^{D-2}]dM/dv$,
which is exactly the same as Eq.~(\ref{T_ab}).
This is the main result of the present paper.

\section{Application to braneworld cosmology}

One possible application of the Vaidya metric in $D=5$ dimensions
is to describe a braneworld that
radiates Kaluza-Klein (KK) gravitons into the bulk.
In such a modeling, the gravitons propagating in the bulk
are approximated by a null fluid and thus are assumed to
be emitted \textit{radially} from the brane.
Due to the emitted gravitons,
the mass of the bulk black hole,
or the dark radiation parameter from the brane point of view,
evolves in time.
A simple radiating braneworld cosmology
in the $Z_2$-symmetric Randall-Sundrum-type setup
has been considered in Refs.~\cite{Langlois:2002ke, Leeper:2003dd}
using the AdS-Vaidya metric~(\ref{AdS-Vaidya}),
and recently the model has been extended to
the asymmetric
case~\cite{Gergely:2004zq, Vernon:2004ae, Jennings:2004wz}~(see
also~\cite{Hebecker:2001nv} and~\cite{Langlois:2003zb}).
However, the dynamics of the dark radiation
has not been discussed so far in the context of
Gauss-Bonnet braneworld
cosmology~\cite{Nojiri:2000gv, Charmousis:2002rc, Germani:2002pt, Lidsey:2002zw, Lidsey:2003sj}.
Since we have obtained the Vaidya-type solution
in Einstein-Gauss-Bonnet gravity in the previous section,
we can now derive an evolution equation
for the dark radiation on the Gauss-Bonnet brane
by making use of our new metric.
From now on we will consider the $D=5$ case with
a negative cosmological constant $\Lambda =-6/\ell^2$.
We will focus ourselves on the ingoing graviton propagation
and hence we take $\epsilon=+1$.


Suppose that in the bulk spacetime~(\ref{Vaidya-GB})
the trajectory of the cosmological brane
is given by $(v(\tau), r(\tau))$.
Here $\tau$ is the proper time on the brane,
so that
\begin{eqnarray}
f\dot v^2-2\dot v \dot r=1,
\end{eqnarray}
where the overdot denotes $d/d\tau$.
The unit tangent to the brane is
$
u^a=(\dot v, \dot r),~u_a=(-f\dot v+\dot r, \dot v),
$
while the unit normal to the brane is
$
n_a=(\dot r,-\dot v),~n^a=(-\dot v, -f\dot v +\dot r).
$
The junction conditions on the brane
under the assumption of a $Z_2$-symmetry
are~\cite{Davis:2002gn, Gravanis:2002wy, Gregory:2003px}
\begin{eqnarray}
K_{ab}=-\frac{\kappa^2}{2}\left(
T_{ab}-\frac{1}{3}Tq_{ab}+\frac{1}{3}\lambda q_{ab}\right)
-2\alpha\left(Q_{ab}-\frac{2}{9}Qq_{ab}\right),
\end{eqnarray}
where
\begin{eqnarray}
Q_{ab}:=2KK_{a\mu}K^{\mu}_{~b}-2K_{a\mu}K^{\mu\nu}K_{\nu b}
+(K_{\mu\nu}K^{\mu\nu}-K^2)K_{ab}+2K R_{ab}
+RK_{ab}+2K^{\mu\nu}R_{a\mu\nu b}-4R_{a\mu}K^{\mu}_{~b},
\end{eqnarray}
and $q_{ab}$ is the induced metric on the brane,
$K_{ab}$ is the extrinsic curvature,
$T_{ab}$ is the energy-momentum tensor of the matter on the brane,
$\lambda$ is a brane tension.
From this junction equations we obtain
the modified Friedmann equation~\cite{Charmousis:2002rc, Maeda:2003vq},
\begin{eqnarray}
\left[H^2+
\frac{1}{4\alpha}\left(1\pm
\sqrt{1-\frac{8\alpha}{\ell^2}+\frac{8\alpha M(\tau)}{r^4}}
\right)
\right]^{1/2}
\left(
\frac{2}{3}\mp\frac{1}{3}
\sqrt{1-\frac{8\alpha}{\ell^2}+\frac{8\alpha M(\tau)}{r^4}}
+\frac{8\alpha}{3} H^2 \right)
=\frac{\kappa^2}{6}(\rho+\lambda),
\label{Friedmann}
\end{eqnarray}
where $r(\tau)$ now plays a roll of the scale factor on the brane and
$H:=\dot r/r$ is the Hubble parameter.
Just for simplicity we have set $k=0$.
In other words, we assume that
the brane universe is spatially flat.
Note that the ``dark radiation'' in the Gauss-Bonnet braneworld
does not behave like a radiation component ($\propto r^{-4}$)
in general.

The local conservation law
in the presence of the bulk matter is given by~\cite{Maeda:2003vq}
\begin{eqnarray}
{\cal D}_bT^b_{~a}=-2{\cal T}_{\mu\nu}q^{\mu}_{~a}n^{\nu},
\label{Bianchi}
\end{eqnarray}
where ${\cal D}_b$ is the covariant derivative with respect to
the brane metric $q_{ab}$.
This takes the same form as
the (non) conservation equation in the Randall-Sundrum
braneworld where Einstein gravity is assumed.
Eq.~(\ref{Bianchi}) relates the energy density on the brane
with the energy flux of the bulk gravitons as
\begin{eqnarray}
\dot{\rho}+3 H(\rho+ p)=-2\sigma,
\label{non-conservation}
\end{eqnarray}
where we have chosen
the ingoing null vector $k^a$ to be normalized so that $k_au^a=1$
and hence $k^r=k_v=1/\dot v$.

Finally, we equate the expressions~(\ref{T_ab}) and (\ref{null_fluid}).
Noting that $\dot v(dM/dv)=\dot M$, we obtain
the evolution equation for the dark radiation parameter:
\begin{eqnarray}
\frac{dM}{d\tau}= \frac{2\kappa^2}{3}\sigma r^4
\left\{
\left[ H^2+\frac{1}{4\alpha}\left(1\pm
\sqrt{1-\frac{8\alpha}{\ell^2}+\frac{8\alpha M(\tau)}{r^4}}
\right)
\right]^{1/2}
-H
\right\}.
\label{DR-evolution}
\end{eqnarray}

In order to solve the coupled
equations~(\ref{Friedmann}),~(\ref{non-conservation}), and~(\ref{DR-evolution}),
we need to specify
the energy flux of the KK gravitons $\sigma$
(and of course the equation of state $p=p(\rho)$).
In the Randall-Sundrum braneworld
one obtains $\sigma \propto \rho^2$
by considering the process
$\psi\bar\psi \to$ KK graviton, where $\psi$
is a particle confined to the brane, and
calculating the scattering amplitude~\cite{Langlois:2002ke}.
Also in the Gauss-Bonnet braneworld
one can in principle do the same procedure to find
$\sigma$ as a function of $\rho$ or the temperature $T$
of the brane particles.
Here we shall present a rather simple argument
to find a rough estimate of the production rate of the bulk gravitons.

The coupling of KK gravitons to the brane matter
is described by the Lagrangian
${\cal L}_{{\rm int}} \sim \kappa u_{\mu\nu}T^{\mu\nu}$,
where $u_{\mu\nu}$ is the canonically normalized gravitational perturbations
evaluated on the brane\footnote{
Strictly speaking, the gravitational coupling to the matter
is shifted due to the Gauss-Bonnet correction~\cite{Neupane:2001st, NandS},
and
the interaction is given by
${\cal L}_{{\rm int}} \sim \kappa (1+4\alpha {\cal K})^{1/2} u_{\mu\nu}T^{\mu\nu}$.
The overall factor $(1+4\alpha {\cal K})$ also appears
in the equation of motion for the gravitational
perturbations.
}, 
and thus the decay late $\Gamma$ of the brane particle is given by
$\Gamma \propto \kappa^2|u_{\mu\nu}|^2$.
Since heavy gravitons with $m\sim T\gg \mu$
mainly contribute to the energy loss from the brane
in the early universe,
where $\mu:=(-{\cal K})^{1/2} \sim \ell^{-1}$ is the bulk curvature scale,
we need to know the behavior of
the mode function $u_{\mu\nu}$ for $m\gg \mu$.
In the Randall-Sundrum braneworld~\cite{RS}
we have $|u_{\mu\nu}|^2\sim$ constant for heavy gravitons,
and hence we obtain $\Gamma\sim T^4/M_5^3$,
where $M_5=\kappa^{-3/2}$ is the fundamental mass scale and
the factor $T^4$ arises for dimensional reason,
leading to $\sigma \sim \Gamma \rho \sim \rho^2 ~(\sim T^8)$.
Unlike the Randall-Sundrum case,
the situation is complicated in the Gauss-Bonnet braneworld
because the graviton mode functions
including Gauss-Bonnet corrections
exhibit a rather nontrivial dependence on $m$
even for $m\gg \mu$~\cite{Neupane:2001st} (Fig.~1).
In fact, assuming that $\alpha \mu^2 \ll 1$,
for the intermediate scale
$\mu \ll T \ll (\alpha\mu)^{-1}$
we have again the Randall-Sundrum behavior
$|u_{\mu\nu}|^2\sim$ constant, so that $\sigma \propto \rho^2$.
At much higher energies $T \gg (\alpha\mu)^{-1}$,
the mode function is suppressed like $|u_{\mu\nu}|^2\sim \mu^2/m^2$.
Consequently we can estimate $\Gamma \sim \mu^2 T^2/M_5^3$,
and thus in this case we find $\sigma \propto \rho^{3/2}$.
These limiting cases are, however, exceptional
and generally it is difficult to give an explicit form of $\sigma$.
For this reason, tackling this issue further seems beyond the scope of this paper.

\begin{figure}[t]
  \begin{center}
    \includegraphics[keepaspectratio=true,height=50mm]{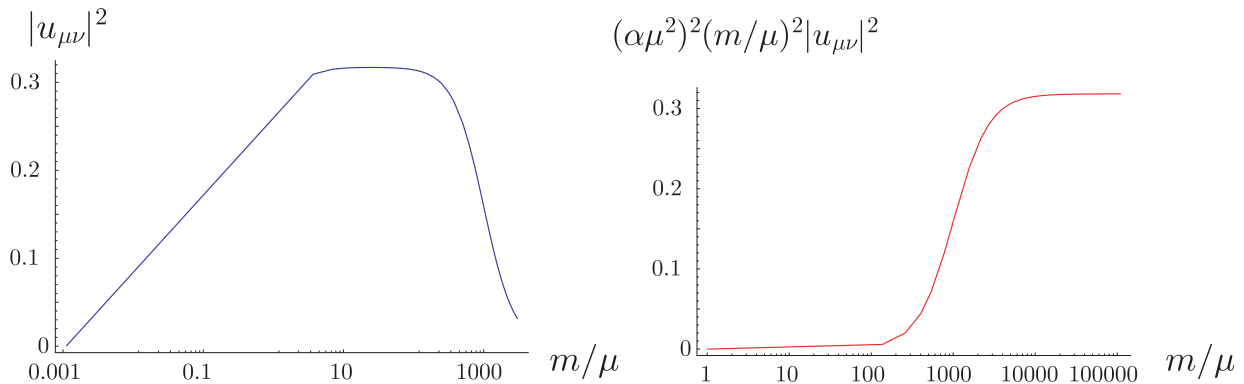}
  \end{center}
  \caption{Mode function evaluated on the brane
  as a function of Kaluza-Klein mass $m$.
  The Gauss-Bonnet coupling is given by $\alpha\mu^2=10^{-3}$.
  (The explicit form of the mode function is found in Ref.~\cite{Neupane:2001st}.)}
  \label{fig:mode.eps}
\end{figure}

\section{Summary}

In the framework of Einstein-Gauss-Bonnet gravity,
which involves a unique combination of quadratic curvature terms,
we have obtained an exact solution that describes
a radiating spacetime filled with a null fluid.
In Einstein gravity such a solution
is called the Vaidya metric, and is constructed by
allowing the mass parameter of
the Schwarzschild or (A)dS-Schwarzschild black hole solution
to depend on the null coordinate: $M\to M(v)$.
Simply and surprisingly, the same procedure
gives the radiating solution of the Einstein-Gauss-Bonnet
field equations with
the energy-momentum tensor of the same null fluid form.
Although the reason is not clear,
this fact
implies that we can generalize
exact solutions in Einstein gravity
to Gauss-Bonnet corrected ones
in a rather na\"{i}ve, straightforward manner.
We hope that our generalized Vaidya solution
enlightens some interesting aspect of Einstein-Gauss-Bonnet gravity.

One of the applications of the solution obtained in this paper 
is using it as a toy model for the Gauss-Bonnet braneworld
that radiates Kaluza-Klein gravitons into the bulk.
We have derived an evolution equation for the ``dark radiation''
coupled to the modified Friedmann equation and
the (non) conservation equation on the brane.

\acknowledgments
T.K. is supported by the JSPS.
He wishes to thank Takahiro Tanaka
for continuous encouragement and
a careful reading of this manuscript.

~\\ {\em Note added:}

After completing this work we became aware of
Ref.~\cite{MaedaH}, which appeared simultaneously to our paper
and deals with the Vaidya-type solution in Einstein-Gauss-Bonnet gravity
in the context of gravitational collapse.



\end{document}